# On the Prediction of the Subharmonic Threshold of an Ultrasound Contrast Agent

Lang Xia

Email: lang4xia@gmail.com

**Abstract**

Subharmonics emitted from microbubble-based ultrasound contrast agents (UCA) are considered as a noninvasive indicator for diagnostic applications in medical fields. Contrast ultrasound imaging using the modality of subharmonics could be also better than that of using harmonics. Unlike the harmonics of an UCA that increase with increasing the amplitude of excitation pressure, subharmonics occur only when the excitation pressure exceeds a threshold value. The excitation pressure inducing the onset of subharmonic components from the UCA during nonlinear oscillations is known as the subharmonic threshold. Although numerous studies on the subharmonics of free bubbles or UCAs have been carried out for decades, several experimental observations—such as the shift of the subharmonic resonance and subharmonic threshold at low excitation pressures—cannot be well characterized by existing theories. An analytical method for resolving these inconsistencies remains incomplete. The present article theoretically investigates the subharmonic threshold of an UCA through the techniques of energy balancing in the phase plane and instability analysis of subharmonic-free oscillations. The influence of shell parameters of the UCA on the subharmonic threshold is studied. Those experimental observations are explained by the amplitude-dependent nonlinear resonance and damping. The effect of nonspherical oscillations on subharmonic emissions is also discussed.

## I. INTRODUCTION

Ultrasound contrast agents (UCAs), made of microscale vesicles or bubbles within gas pockets being encapsulated, can significantly increase the contrast-to-tissue ratio in biomedical imaging (De Jong, Emmer, Van Wamel, & Versluis, 2009; Karandish et al., 2018; Kulkarni et al., 2018; Stride et al., 2020; Xia et al., 2017). This capacity is mainly due to the oscillatory motions of UCAs subjected to excitation pressures. Depending on the excitation amplitudes UCAs may oscillate linearly or nonlinearly and exhibit rich dynamic behaviors, which have been studied via various modifications of a Rayleigh-Plesset equation (Church, 1995; Doinikov & Dayton, 2007; Marmottant et al., 2005; Mobadersany & Sarkar, 2019). Although linear oscillations of UCAs are preferred in the diagnostic applications for potential safety concerns, they do not always render sufficient quality and clarity to ultrasonic imaging reconstruction because of the interferences from linear tissue signals or echoes. Ultrasonic signals emitted from nonlinear oscillations of UCAs are then employed to overcome the limitation associated with using linear ultrasound because in the frequency domain they generate new nonlinear components that are not present in linear ultrasound. Image reconstruction using these nonlinear components may not suffer from the interferences (de Jong, Bouakaz, & Frinking, 2000; Shankar, Krishna, & Newhouse, 1998).

The nonlinear oscillation of an UCA being driven with a sinusoidal signal is usually characterized by the generation of harmonics, which are integer multiples of the excitation frequency. Under proper conditions, however, the UCA in an ultrasound field will oscillate periodically with frequencies equal to fractions of the excitation frequency, of which the one-half is often the strongest and simply referred to the subharmonic of the UCA. A unique feature of subharmonics of free bubbles or UCAs is that their presence requires the amplitude of the excitation pressure to exceed a threshold value. Subharmonic signals were first discovered in cavitating liquids, of which the threshold was thought to correlate with the onset of transient cavitation (Mellen, 1954). Later, an experimental investigation of single free bubbles suggested that the subharmonic signals could exist in liquids without any transient cavitation and exhibited threshold behavior (Neppiras, 1969). It also indicated that the surface instability of big bubbles could link with the subharmonic threshold (D. Hsieh, 1974). Now the generation of subharmonic signals in those early acoustic experiments can be explained theoretically by the instability of nonlinear oscillations in which the damping plays a key role. Due to the thresholding feature, subharmonic signals of UCAs have been proposed as an effective and noninvasive estimator to measure organ-level blood pressures in medical fields (Shi, Forsberg, Raichlen, Needleman, & Goldberg, 1999).

One of the first theoretical work for predicting the subharmonic threshold of a free bubble considering the effects of damping was done by Eller and Flynn (Eller & Flynn, 1969). They considered the perturbations about the linear bubble response and then obtained a stability equation of the Hill's type, of which the onset of the instability leads to the generation of the subharmonic oscillations. The damping of the bubble opposed the generation of the subharmonic and built the threshold. Their results showed that the subharmonic threshold was minimal when the excitation frequency was about two times the linear resonance frequency of the bubble, whereas the bubble without damping did not have such a threshold. Prosperetti later on expanded a Rayleigh-Plesset equation to the third order and subsequently obtained an analytical solution for subharmonic amplitude using the Bogolyubov-Krylov asymptotic method (Prosperetti, 1974). These asymptotic results were then compared with numerical simulations, which showed very satisfactory agreements near the subharmonic resonance (Lauterborn, 1976; Prosperetti, 1977). Hsieh introduced a method of calculus of variation to predict the subharmonic threshold of a bubble occurring at the excitation frequency twice of the linear resonance (D. Y. Hsieh, 1975). Although this method is simple in concept and mathematically more general, it received less attention because of involved algebraic calculations in the derivations.

When it comes to the theoretical studies of subharmonic thresholds of UCAs (or coated bubbles), the situation becomes more complicated. First, the shell of an UCA introduces additional nonlinear terms in a Rayleigh-Plesset equation, making utilization of the aforementioned techniques more difficult. Second, UCAs with different shell materials (e.g., polymer, protein, and lipids) require different types of Rayleigh-Plesset equations, bringing significant complexity in developing a unified method for predicting subharmonic thresholds. Following the approach that was used in (Prosperetti, 1974), analytical results of subharmonic thresholds for UCAs of various shells were obtained (Kimmel, Krasovitski, Hoogi, Razansky, & Adam, 2007; Shankar, Krishna, & Newhouse, 1999; Sijl et al., 2010). These results evaluated thresholds for both the existence of the subharmonic and the onset of subharmonic instability, of which the latter is commonly mentioned as the subharmonic threshold in the literature and is the focus of this article. Comparisons of perturbative and numerical results were not given in the above-mentioned references. A general derivation focusing on the existence of subharmonics of UCAs coated with different shells was attempted by (Prosperetti, 2013), of which the numerical comparisons suggest that further developments are still needed to accurately predict the threshold for the subharmonic instability.

In addition to the conventional perturbative methods, periodic oscillations of a nonlinear oscillator have been studied extensively by phase plane techniques, of which the existence of subharmonic oscillation is predicted by the simple zeros of a subharmonic Melnikov's function (Guckenheimer & Holmes, 2013). This analytical method investigates the geometric behaviors of a forced and damped oscillator (also known as a perturbed system) around a closed orbit of the same oscillator but neglects the forcing and damping terms (namely, the unperturbed system, in which the governing equation is autonomous) in a phase plane. It then utilizes regular perturbation to derive the subharmonic Melnikov's function that only involves the solutions of the autonomous equation, greatly simplifying the mathematical computation. Although the Melnikov's method is well-established and commonly seen in studies dealing with weakly damped and forced nonlinear oscillators, it is rarely applied to analyze chaotic dynamics and subharmonics of bubbles (Smereka, Birnir, & Banerjee, 1987), possibly due to its elusive physical significance. An equivalent approach but with more physical insights was then developed and used to predict the chaotic motion and subharmonic threshold of certain nonlinear oscillators (Dankowicz, 1996; Kovacic & Brennan, 2011).

Almost all the results from the regular perturbative analysis convey a message that the minimum subharmonic threshold of an UCA occurs only when the driven frequency is twice the linear resonance of the UCA (also known as the subharmonic resonance). Recent numerical results, however, show that the minimum threshold for subharmonic generation from an UCA moves around twice the linear resonance frequency (Jimenez-Fernandez, 2018; Katiyar & Sarkar, 2012). If the amplitude of excitation pressure is decreased, then the minimum subharmonic threshold approaches twice the linear resonance frequency. The shift of the minimum subharmonic threshold has not been explained from the theoretical point of view regardless of the above-mentioned analytical works. Several experimental observations--which the minimum subharmonic threshold could occur near the main resonance and be significantly smaller than that of theoretical predictions--have not been interpreted in detail either (Chomas, Dayton, May, & Ferrara, 2002; Faez, Skachkov, Versluis, Kooiman, & de Jong, 2012; Lotsberg, Hovem, & Aksum, 1996; Shankar et al., 1999). Furthermore, the influence of nonspherical oscillations of an UCA on its subharmonic threshold remains undiscussed.

Differently from other studies, this article first analyzes the energy form of a modification of a Rayleigh-Plesset equation for an UCA, of which the unperturbed system is solved analytically. The exact solution of the unperturbed system is transformed into the nonlinear resonance frequency of the UCA, which enables deriving a subharmonic Melnikov's function. Forcing this function to zero yields an equation for predicting the existence of the subharmonic component of the UCA. Then by separating the oscillation of the UCA into subharmonic-free and subharmonic components, a relatively general analysis of the instability of the governing equation is performed. An explicit equation for predicting the threshold of subharmonic instability is then derived. Several nonlinear phenomena related to the subharmonic of the UCA, e.g., shifting of subharmonic threshold, are explained based on the analytical results.

## II. MATHEMATICAL ANALYSIS

We use an UCA of the viscoelastic shell model, which represents a major UCA type nowadays, to perform the mathematical derivation.

### A. Radial oscillation

A Rayleigh-Plesset type equation for the spherical dynamics of an UCA coated with a viscoelastic shell oscillating in a compressible liquid can be written in the form (Marmottant et al., 2005)

$$\rho\left(R\ddot{R}+\frac{3}{2}\dot{R}^2\right)=P_{g0}\left(\frac{R_0}{R}\right)^{3\kappa}\left(1-\frac{3\kappa\dot{R}}{c_0}\right)-4\kappa^s\frac{\dot{R}}{R^2}-2E^s\frac{1}{R}\left(\frac{R^2}{R_0^2}-1\right)-2\gamma_0\frac{1}{R}-4\mu\frac{\dot{R}}{R}-p_0+p(t) \quad (1)$$

where $\rho$ is the density of the surrounding liquid, $R$ is the instantaneous radius of the bubble around the equilibrium radius $R_0$, and $\dot{R}=dR/dt$, $\ddot{R}=d^2R/dt^2$, $P_{g0}$ is the initial pressure inside the bubble, $\kappa$ is the polytropic constant, $c_0$ is the sound speed in the host liquid without UCAs, $\kappa^s$ and $E^s$ are the dilatational viscosity and elasticity of the UCA shell, respectively, $\mu$ is the viscosity of the host liquid, $\gamma_0$ is a reference value of the surface tension, $p_0$ is the hydrostatic pressure in the liquid, and $p(t)=P_A f(t)$ is the excitation pressure with $P_A$ the amplitude of the periodic function $f(t)$. In the above equation we have neglected the effects of gas diffusion and assumed the elasticity to be a constant (Xia, 2019).

We first partly nondimensionalize the above equation by scaling the radius and time as $R=R_0 r$ and $t=\tau/\omega_0$, respectively. Upon multiplying both sides of the nondimensionalized equation with $r^2 dr/d\tau$, we obtain

$$\frac{d}{d\tau}(\mathcal{T}+\mathcal{V})+\frac{d}{d\tau}\mathcal{D}=\frac{d}{d\tau}\mathcal{W} \quad (2)$$

with

$$\begin{aligned}
\mathcal{T} &= \frac{1}{2}r^3\left(\frac{dr}{d\tau}\right)^2 \\
\mathcal{V} &= \frac{1}{\rho R_0^2 \omega_0^2}\left[\frac{P_{g0}}{3\kappa-3}\frac{1}{r^{3\kappa-3}}-\frac{1}{R_0}(E^s-\gamma_0)r^2+\frac{1}{3}p_0 r^3+\frac{E^s}{2R_0}r^4\right] \\
\frac{d\mathcal{D}}{d\tau} &= \frac{1}{\rho R_0^2 \omega_0}\left[P_{g0}\frac{3\kappa R_0}{c_0}\frac{1}{r^{3\kappa}}+\frac{4\kappa^s}{R_0}\frac{1}{r^2}+4\mu\frac{1}{r}\right]\left(r\frac{dr}{d\tau}\right)^2 \\
\frac{d\mathcal{W}}{d\tau} &= \frac{P_A}{\rho R_0^2 \omega_0^2}f(\frac{\omega}{\omega_0}\tau)r^2\frac{dr}{d\tau}
\end{aligned} \quad (3)$$

Here $\mathcal{T}$ is the kinetic energy that relates to the inertial force, $\mathcal{V}$ is the potential energy that relates to the restoring force. The dissipation energy induced by the damping and the work done by the external force are represented by $\mathcal{D}$ and $\mathcal{W}$, respectively. Equations in (3) shows that the viscoelastic shell contributes to the potential energy through the dilatational elasticity and the dissipation energy through the dilatational viscosity. Equation (2) is also called a perturbed system. If the damping and external force are absent in Eq.(2), it will be reduced to the so-called unperturbed system. Due to the presence of the damping and external force terms, no exact solution of Eq.(2) has been found by far (Xia, 2018).

## B. Exact solution and nonlinear resonance of the unperturbed system

Equation (2) reduces to an autonomous differential equation (conservative system) of the unperturbed system when the damping and external force are neglected, of which the first integral is in the form of

$$\mathcal{T} + \mathcal{V} = \mathcal{H} \tag{4}$$

where $\mathcal{H}$ is an integral constant that is related to the total energy and is determined by the initial conditions. If Eq.(4) has periodic solutions, $\mathcal{H}$ will determine a set of loops in the phase plane. By using the initial conditions $r(0) = 1$ and $dr(0)/d\tau = 0$, Eq.(4) can be further integrated and solved for the dimensionless time, which has the exact form

$$\tau = \frac{1}{\sqrt{2}} \int_1^r \frac{a^{3/2}}{[\mathcal{H} - \mathcal{V}(a)]^{1/2}} da \tag{5}$$

where

$$\mathcal{H} = \frac{1}{\rho R_0^2 \omega_0^2} \left[ \frac{\kappa}{3(\kappa-1)} p_0 + \frac{3\kappa-1}{3(\kappa-1)} \frac{\gamma_0}{R_0} - \frac{E^s}{2R_0} \right] \tag{6}$$

We remark that of the two possible roots from solving Eq.(4) for $\tau$, we keep the positive one for the positive time. Although Eq.(5) cannot generally be integrated into an elementary function, it is an analytical expression for the time and is important in deriving the threshold for the existence of the subharmonic orbit.

When a periodic solution exists, the period of this unperturbed system can be written as

$$T = \frac{2\sqrt{2}}{\omega_0} \int_1^A \frac{r^{3/2}}{[\mathcal{H} - \mathcal{V}(r)]^{1/2}} dr \tag{7}$$

where $A$ is the maximal amplitude of the periodic motion. Finally, we can compute the nonlinear resonance frequency of the unperturbed system via $\Omega = 2\pi/T$, or

$$\Omega = \frac{\pi \omega_0}{\sqrt{2} \int_1^A \frac{r^{3/2}}{[\mathcal{H} - \mathcal{V}(r)]^{1/2}} dr} \tag{8}$$

.

## C. Threshold for the existence of subharmonic oscillation

Qualitatively, the subharmonic orbit (of 1/2 order) of a perturbed system favors a condition that the nonlinear restoring force contains terms of the highest degree of the power not less than 2 (Hayashi, 2014). For the present case, $d\mathcal{V}/d\tau$ (the nonlinear restoring force) in Eq.(2) always has such terms, suggesting the subharmonic oscillation could exist in this system. The traditional Melnikov's method can quantitively predict the existence of the subharmonic solution of the perturbed system [Eq.(2)] via the solution of the unperturbed system [Eq.(5)]. It is often convenient to convert the unperturbed system to a Hamiltonian system for directly utilizing the subharmonic Melnikov's function. This can be done by substituting $R = R_0 x^{2/5}$ into Eq.(1), that is

$$\begin{cases} \dot{x} = u \\ \dot{u} = \dfrac{5}{2} \dfrac{1}{\rho R_0^2 \omega_0^2} \left( p_0 x^{1/5} - P_{g0} \dfrac{1}{x^{(6\kappa-1)/5}} + \dfrac{2E^s}{R_0} x^{3/5} - \dfrac{2}{R_0} \left( E^s - \gamma_0 \right) \dfrac{1}{x^{1/5}} \right) \\ \quad + \left[ \dfrac{1}{\rho R_0^2 \omega_0} \left( P_{g0} \dfrac{3\kappa R_0}{c_0} \dfrac{1}{x^{(6\kappa+2)/5}} + \dfrac{4\kappa^s}{R_0} \dfrac{1}{x^{6/5}} + 4\mu \dfrac{1}{x^{4/5}} \right) \dot{x} - \dfrac{5}{2} \dfrac{P_A}{\rho R_0^2 \omega_0^2} f(\dfrac{\tau}{\omega_0}) x^{1/5} \right] \end{cases} \quad (9)$$

If the contribution from the external force and damping is small [the terms in the last square bracket in Eq.(9)], the existence of the subharmonic orbit of the perturbed system Eq.(2) can be estimated from the solution of the corresponding unperturbed system. We thus can define the subharmonic Melnikov's function $M$ as follows [see page 195 in (Guckenheimer & Holmes, 2013)]

$$M = \dfrac{1}{\rho R_0^2 \omega_0^2} \int_0^{2\tilde{T}} \dot{x} \left[ \dfrac{5}{2} P_A f(\dfrac{\tau}{\omega_0}) x^{1/5} - \omega_0 \left( P_{g0} \dfrac{3\kappa R_0}{c_0} \dfrac{1}{x^{(6\kappa+2)/5}} + \dfrac{4\kappa^s}{R_0} \dfrac{1}{x^{6/5}} + 4\mu \dfrac{1}{x^{4/5}} \right) \dot{x} \right] d\tau \quad (10)$$

or change back to

$$M = \dfrac{25}{4} \dfrac{1}{\rho R_0^2 \omega_0^2} \int_0^{2\tilde{T}} \dot{r} \left[ P_A f(\dfrac{\tau}{\omega_0}) r^2 - \omega_0 \left( P_{g0} \dfrac{3\kappa R_0}{c_0} \dfrac{1}{r^{3\kappa}} + \dfrac{4\kappa^s}{R_0} \dfrac{1}{r^2} + 4\mu \dfrac{1}{r} \right) r^2 \dot{r} \right] d\tau \quad (11)$$

where $\tilde{T} = 2\pi\omega_0 / \omega$, and $\omega$ is the frequency of the external force $f$. Forcing $M = 0$ gives the critical values of the subharmonic existence or the threshold of subharmonic existence. The traditional subharmonic Melnikov's method analyzes the dynamic behaviors of the perturbed system around the periodic orbit from a geometric view by assuming that the integrand in the above equation is sufficiently small. Alternatively, we can derive the subharmonic Melnikov's function purely from the aspect of energy balance, which has more physical insights. For the perturbed system, we may still denote the energy by $\mathcal{H}$, but note that it is a time-dependent variable now. We then define

$$\dfrac{d}{d\tau}\mathcal{H} := \dfrac{d}{d\tau}(\mathcal{T} + \mathcal{V}) = \dfrac{d}{d\tau}\mathcal{W} - \dfrac{d}{d\tau}\mathcal{D} \quad (12)$$

as the rate of change in energy. If the energy dissipates more than the work done by the external force, the oscillation will end up at an equilibrium point. On the other hand, if the energy dissipates less than the work done by the external force, the oscillation may go unstable. Integration of the above equation over the period of $2\tilde{T}$ yields

$$\mathcal{H}(2\tilde{T}) - \mathcal{H}(0) = \int_0^{2\tilde{T}} \left( \dfrac{d}{d\tau}\mathcal{W} - \dfrac{d}{d\tau}\mathcal{D} \right) d\tau \quad (13)$$

If the motion of the perturbed system possessing the periodic orbit is stable, $\mathcal{H}$ will return to its original value after $2\tilde{T}$, and then we obtain

$$\int_0^{2\tilde{T}} \left( \frac{d}{d\tau}\mathcal{W} - \frac{d}{d\tau}\mathcal{D} \right) d\tau = 0 \tag{14}$$

on the subharmonic orbit of the perturbed system. Eq.(14) is an exact relation for any system possessing a stable subharmonic orbit. However, obtaining an explicit expression for Eq.(14) requires exact solutions of Eq.(2), which have not been found in the literature by far. Noticing that the exact solution of the unperturbed system has been given by Eq.(5), we thus may be able to utilize this solution after approximating the perturbed system about the periodic orbit of the unperturbed system. When the unperturbed solution is involved, this method of energy balance is essentially the same as that of the foregoing Melnikov's method (Dankowicz, 1996), and Eq.(14) is also equivalent to finding simple zeros of Eq.(11). With the assistance of Eq.(5), we finally have the following relationship for the excitation pressure and frequency

$$P_A = \frac{\int_{r_1}^{r_2} \omega_0 \left[ P_{g0} \frac{3\kappa R_0}{c_0} \frac{1}{r^{3\kappa}} + \frac{4\kappa^s}{R_0} \frac{1}{r^2} + 4\mu \frac{1}{r} \right] \left[ 2r\mathcal{H} - 2r\mathcal{V}(r) \right]^{1/2} dr}{\int_{r_1}^{r_2} f\left( \frac{\tau}{\omega_0} \right) r^2 dr} \tag{15}$$

where $r_1$ and $r_2$ are obtained by solving

$$\begin{aligned} 0 &= \frac{1}{\sqrt{2}} \int_1^{r_1} \frac{a^{3/2}}{\left[ \mathcal{H} - \mathcal{V}(a) \right]^{1/2}} da \\ 2\tilde{T} &= \frac{1}{\sqrt{2}} \int_1^{r_2} \frac{a^{3/2}}{\left[ \mathcal{H} - \mathcal{V}(a) \right]^{1/2}} da \end{aligned} \tag{16}$$

The periodic function is usually of the form $f(\tau/\omega_0) = \cos(\omega\tau/\omega_0)$. Equation (15) is the threshold for the existence of the subharmonic orbit, in which the first square bracket of the integrand in the numerator contains exactly the damping terms that are in the nondimensionalized Eq.(1). It is not difficult to see that these damping effects oppose the generation of the subharmonic and build up the threshold. The threshold value will be zero if the damping terms are neglected.

**D. Threshold for the instability of subharmonic oscillation**

The threshold given by Eq.(15) is a necessary but not sufficient condition for the occurrence of the subharmonic of an UCA. Depending on the initial conditions of the UCA motion, the subharmonic component may or may not exist. Studying the instability of the subharmonic orbit would provide the sufficient condition for the occurrence of the subharmonic. In the method of phase plane (Melnikov's analysis), this proceeds with the symplectic transformation of the unperturbed system to action-angle coordinates. However, Melnikov's analysis does not guarantee the accuracy of the prediction, particularly when the dissipation is large and time-dependent (Guckenheimer & Holmes, 2013).

On the other hand, we see that the occurrence of the subharmonic requires the subharmonic-free oscillations to be unstable, which suggests that investigation of the stability of periodic oscillations may also predict the threshold for the instability of subharmonic oscillation. Thus, we can assume $r = X + y$, where $X$ is the subharmonic-free periodic oscillations, and $y$ is the subharmonic oscillation. Substitution of this relation into the governing equation will give a differential equation with respect to the variable $y$, with $X$ contributing to the time-dependent coefficients. To study the threshold value, we may assume the subharmonic component to be sufficiently small ( $y \ll X$ ) and then keep only the first-order term of $y$. Thus, we will obtain a linear differential equation for $y$. The accuracy of this differential equation is determined by the subharmonic-free solution $X$, which can be expressed by the Fourier series. To illustrate this idea, we also truncate the $X$ to the first order and obtain the following stability equation

$$\left[1+(3\kappa+1)X\right]\ddot{y} + \left[\omega_0\delta_0 + 3\dot{X} + \omega_1\delta_1 X\right]\dot{y}$$
$$+ \left[\omega_0^2 + (3\kappa+1)\ddot{X} + \omega_1\delta_1\dot{X} + \omega_1^2 X - \frac{3\kappa}{\rho R_0^2}P_A\cos(\omega t)\right]y = \frac{1}{\rho R_0^2}P_{g0} \qquad (17)$$

with the coefficients defined by

$$\omega_0^2 = \frac{1}{\rho R_0^2}\left(3\kappa p_0 + \frac{4E^s}{R_0} + (3\kappa-1)\frac{2\gamma_0}{R_0}\right),$$
$$\delta_0 = \frac{4}{\rho R_0^2 \omega_0}\left(\frac{3\kappa R_0}{4c_0}P_{g0} + \frac{\kappa^s}{R_0} + \mu\right) \qquad (18)$$

and

$$\omega_1^2 = \frac{1}{\rho R_0^2}\left(3\kappa(3\kappa-1)p_0 + 3(4\kappa-1)\frac{2E^s}{R_0} + (3\kappa-2)(3\kappa-1)\frac{2\gamma_0}{R_0}\right),$$
$$\delta_1 = \frac{4}{\rho R_0^2 \omega_1}\left((3\kappa-2)\frac{\kappa^s}{R_0} + (3\kappa-1)\mu\right) \qquad (19)$$

of which the equations in (18) are the linear resonance and linear damping constant (Xia, 2020), respectively. The subharmonic threshold thus can be derived when the subharmonic solution of Eq.(17) exists. Following the first-order approximation, we assume

$$X = \frac{1}{2}A\exp(\mathbf{i}\omega t) + c.c.$$
$$y = \frac{1}{2}C\exp(\frac{1}{2}\mathbf{i}\omega t) + c.c. \qquad (20)$$

where $A$ is the complex amplitude of the fundamental oscillation (Xia, 2020), and $C$ is the complex amplitude of the subharmonic oscillation. Here $c.c.$ is short for complex conjugate. Substitution of Eq.(20) into Eq.(17) with keeping only subharmonic frequency components yields

$$\left[-\frac{1}{4}\omega^2 + \frac{1}{2}i\omega\omega_0\delta_0 + \omega_0^2\right]C = \frac{1}{2}\frac{P_A}{\rho R_0^2}\left[-\frac{-\frac{15\kappa-1}{4}\omega^2 + \frac{1}{2}i\omega\omega_1\delta_1 + \omega_1^2}{-\omega^2 + i\omega\omega_0\delta_0 + \omega_0^2} + 3\kappa\right]\bar{C} \quad (21)$$

Here the bar over *C* stands for the complex conjugate. A nontrivial solution of *C* in the above equation requires the determinant vanished, that's

$$P_A = 2\rho R_0^2 \frac{\left|-\frac{1}{4}\omega^2 + \frac{1}{2}i\omega\omega_0\delta_0 + \omega_0^2\right| \cdot \left|-\omega^2 + i\omega\omega_0\delta_0 + \omega_0^2\right|}{\left|\frac{3\kappa-1}{4}\omega^2 + i\omega\left(3\kappa\omega_0\delta_0 - \frac{1}{2}\omega_1\delta_1\right) + 3\kappa\omega_0^2 - \omega_1^2\right|} \quad (22)$$

which is analog to the transition curve defined in the theory of Mathieu's equation and determines the instability region (also called 'tongue') in the frequency-amplitude plane (Kovacic, Rand, & Mohamed Sah, 2018). Inside the tongue, the solution is unstable. Therefore, Eq.(22) characterize the threshold behavior for the instability of the subharmonic oscillation. Note that the very similar expression was derived by Prosperetti, except for that he deemed it insufficient in determining the instability of the subharmonic solution (Prosperetti, 2013).

## III. RESULTS AND DISCUSSIONS

### A. Numerical Comparisons

#### 1. Free bubble (UCA without coating)

To compare the analytical solutions presented in this article with the numerical results obtained for free bubbles (Lauterborn, 1976; Prosperetti, 1977), we take the physical constants from these two references. The host medium is the water with: density $\rho = 1000 \text{ Kg/m}^3$, viscosity $\mu = 0.001 \text{ N} \cdot \text{s/m}^2$, hydrostatic pressure $p_0 = 101325 \text{ Pa}$. The sound speed is assumed infinity to conform with the numerical results simulated in the incompressible water. The initial radius of the bubble is $R_0 = 1 \times 10^{-5} \text{ m}$, with the polytropic constant $\kappa = 1.33$ and surface tension $\sigma_0 = 0.072 \text{ N/m}$. Additionally, both the shell dilatational viscosity $\kappa^s$ and elasticity $E^s$ are taken zero for the free bubble. *Figure 1* lists the thresholds for the existence (green-dash curve) and the instability (blue-solid curve) of the subharmonic calculated theoretically from Eq.(15) and Eq.(22), respectively, along with the numerical results represented by the dark circles. As Eq.(15) predicts only the possible region (above the curve) where the subharmonic may exist, it provides a lower bound for the threshold. In contrast, Eq.(22) matches the numerical results very well. Considering only the first-order approximation employed here, this explicit expression is quite effective for the prediction of the subharmonic threshold of the free bubble.

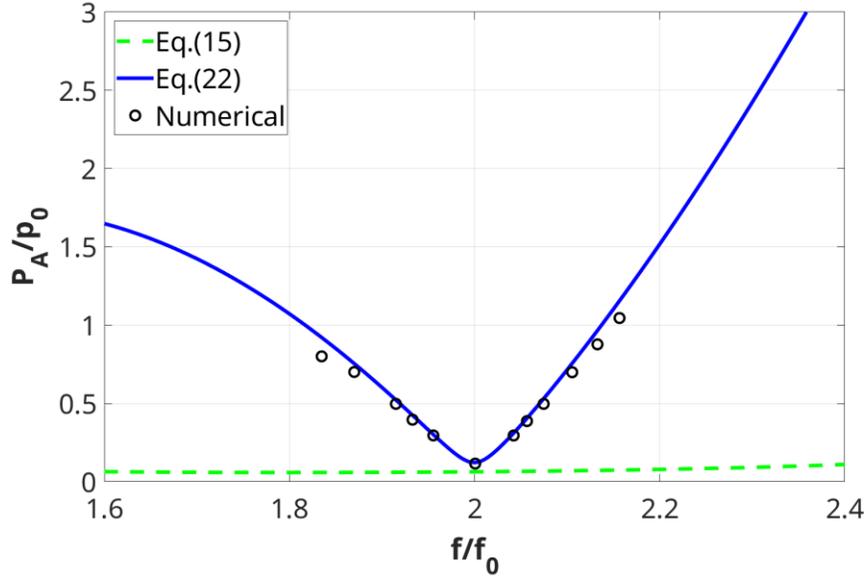

**Figure 1**: Comparisons between Eq.(15), Eq.(22), and Lauterborn's numerical results (free bubbles).

## 2. UCA

The comparison is also carried out on an UCA (coated-lipid microbubble). The physical constants of water (host medium) used in the numerical simulations are listed as follows: density $\rho = 1000 \, \text{Kg/m}^3$, hydrostatic pressure $p_0 = 101325 \, \text{Pa}$, sound speed $c_0 = 1500 \, \text{m/s}$, viscosity $\mu = 0.002 \, \text{N} \cdot \text{s/m}^2$ (is doubled to account for thermal effect). The physical constants of the UCA are: initial radius $R_0 = 2.5 \times 10^{-6} \, \text{m}$, polytropic constant $\kappa = 1.07$, shell dilatational viscosity $\kappa^s = 5 \times 10^{-9} \, \text{N} \cdot \text{s/m}$, shell elasticity $E^s = 0.1 \, \text{N/m}$, surface tension of the encapsulation $\sigma_0 = 0 \, \text{N/m}$. These shell parameters were estimated from lipid-coated microbubbles (Xia, Porter, & Sarkar, 2015). The subharmonic threshold of numerical results is obtained from directly solving Eq.(1) followed by checking the subharmonic components of the radius curve [numerical solution of Eq.(1)] in the frequency domain (caution should be made to avoid the transient part of the curve, as well as spectra leaking due to fast Fourier transform). Similar to the free bubble, the dashed curve of Eq.(15) does not meet the curve of Eq.(22) in *Figure 2*. Note that the curve of the existence of the subharmonic usually touches the curve of the instability of the subharmonic at $f = 2f_0$ (Prosperetti, 1977; Shankar et al., 1999). Here we don't observe the contact of the two curves, probably due to the limitation of Melnikov's analysis. Because the Melnikov's subharmonic theorem requires that the dissipation terms in the perturbed system are linear and sufficiently small. As for the threshold of subharmonic instability, the curve of Eq.(22) and the numerical results match well near $f/f_0 = 2$, spanning from 1.95 to 2.05. Outside this region, the analytical curve deviated from the numerical results. For the frequency ratio less than 2, threshold values predicted by Eq.(22) are always lower than the numerical values, while the analytical threshold values get larger than the numerical results for $f/f_0 > 2$. The lower frequency ($f/f_0 < 2$) deviation is mainly due to the first order approximation of the X that neglected higher order harmonics, and the higher frequency deviation could be caused by the first order approximation of *Y* that neglects higher order subharmonics. Nevertheless, Eq.(22) captures the essential features of the subharmonic threshold when the excitation frequencies are far away from twice the linear resonance frequency of the UCA.

In the following sections, we will mainly focus on the threshold of subharmonic instability, and hereafter, without explicitly expressing, the subharmonic threshold always refers to Eq.(22).

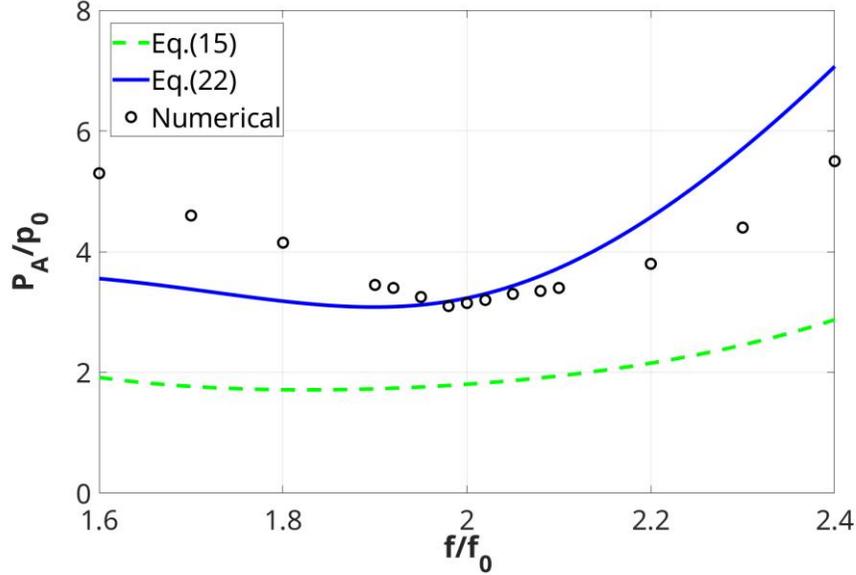

**Figure 2**: Comparisons between Eq.(15), Eq.(22), and the numerical results obtained directly from Eq.(1) (UCA).

**B. Influence of encapsulation parameters**

The dilatational viscosity $\kappa^s$ and elasticity $E^s$ are two major shell parameters that affect the dynamic behaviors of the UCA with a lipid coating. $\kappa^s$, contributing to the overall damping of the UCA, increases the subharmonic threshold, of which the influence is demonstrated in ***Figure 3***(a). In computing these curves via Eq.(22), the elasticity is fixed at 0.5 N/m, while the dilatational viscosity varies from 0 to $10^{-8}$ N·s/m, as indicated by the dashed arrow. Besides the increased threshold with increasing $\kappa^s$, the minimum subharmonic threshold, starting from almost *f/f₀* = 1.98, also goes to lower frequencies. It suggests that the damping contributes to the shift of the subharmonic resonance that was mentioned in the Introduction. For the reference, the curve of the subharmonic threshold without damping is plotted in the blue dashed curve in ***Figure 3*** (a). The effect of the dilatational elasticity $E^s$ is investigated by varying it from 0 to 1 N/m while keeping $\kappa^s = 5 \times 10^{-9}$ N·s/m. The results are illustrated by the family of blue curves in ***Figure 3*** (b), which suggests that the increased dilatational elasticity helps build the subharmonic threshold as well. This is because an UCA with larger elasticity tends to be 'stiffer', requiring higher excitation pressures for achieving prominent nonlinear oscillations that trigger a potential subharmonic oscillation. Unlike the effect of the dilatational viscosity that pushes the minimum subharmonic threshold to lower frequencies when it increases, the dilatational elasticity pulls the threshold back to twice the linear resonance, from *f/f₀* = 1.76 in the bottom curve to almost 2 in the top curve. Therefore, while both parameters can raise the minimum subharmonic threshold, they shift it in opposite directions.

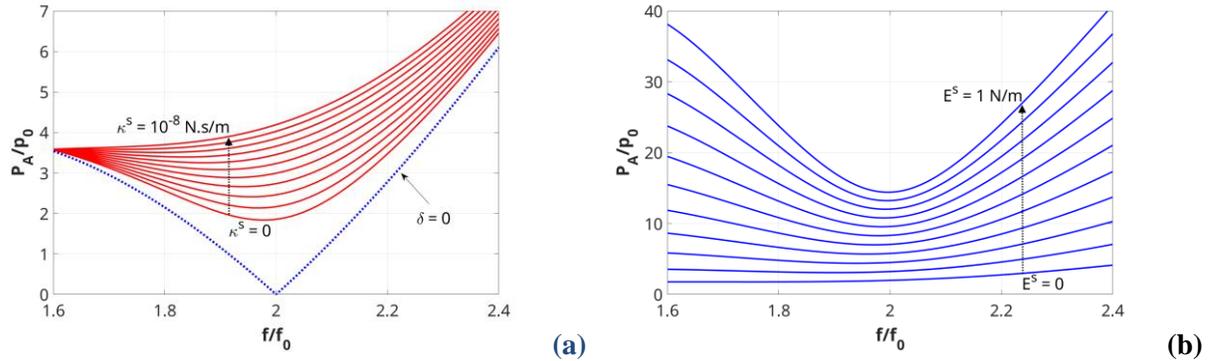

**Figure 3**: The subharmonic threshold vs excitation frequency with varying the dilatational viscosity while the value of elasticity is fixed at 0.5 N/m (a), and with varying the dilatational elasticity while the value of the viscosity is fixed at 5×10$^{-9}$ N·s/m (b). Here $R_0 = 2.5 \times 10^{-6}$ m.

## C. Influence of polytropic constant and UCA's initial radius

In the preceding section we saw that the subharmonic threshold increases progressively with increasing the dilatational viscosity or elasticity as each of the shell parameters contributes solely to either the linear resonance frequency or damping [see equations in (18)]. However, the influence of the polytropic constant or UCA radius on the subharmonic threshold cannot be easily derived from these equations because of their simultaneous appearances in each of Eq. (18). We thus again visualize Eq.(22). **Figure 4** presents a family of the frequency-dependent threshold curves in the range of 1.6 to 2.4 folds of the resonance frequency $f_0$, each of which is calculated with a specific polytropic constant starting from $\kappa^s = 1$ and ending with $\kappa^s = 1.4$. It shows that, for this set of shell parameters, the subharmonic threshold decreases gradually with increasing the polytropic constant, but the change is not prominent. In the contract, the variation of the threshold curve with respect to the UCA's initial radius is more significant. **Figure 5** demonstrates that an UCA of radius 1 µm has a significantly larger threshold as indicated by the top curve. When the radius is increased to 2 µm, the threshold drops drastically (see the one below the top curve). The speed of the drop decreases with further increasing the radius. It implies that a smaller UCA could possess a significantly higher subharmonic threshold than those with the same shell parameters.

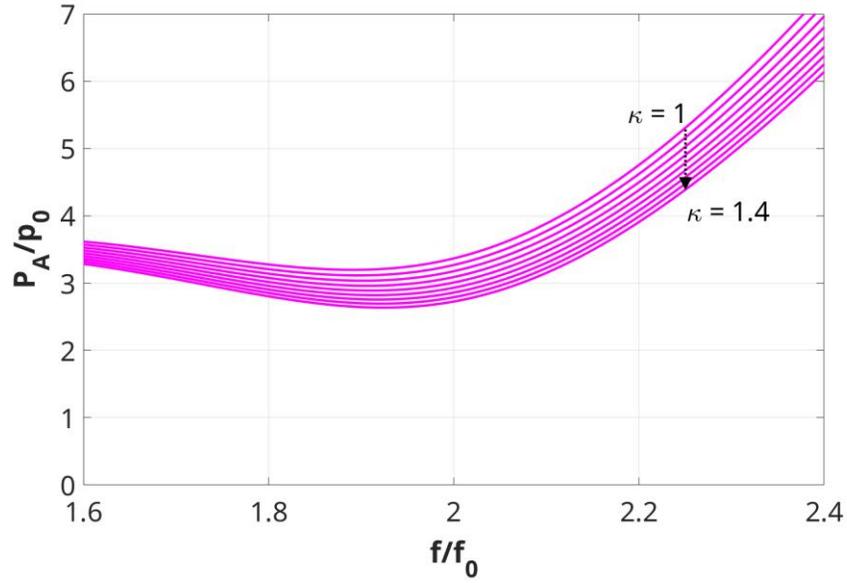

**Figure 4**: The subharmonic threshold vs excitation frequency with varying the polytropic constant. Here $R_0 = 2.5 \times 10^{-6}$ m, $\kappa^s = 5 \times 10^{-9}$ N·s/m, $E^s = 0.1$ N/m.

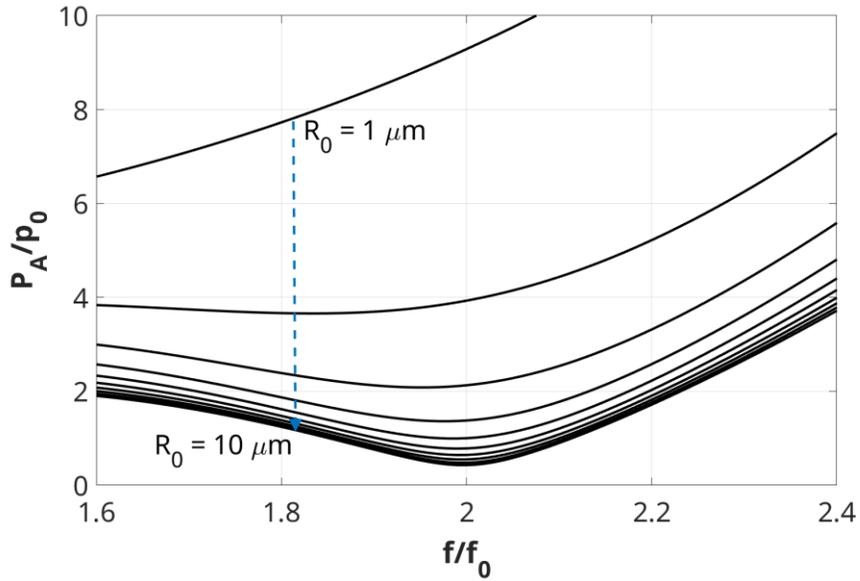

**Figure 5**: The subharmonic threshold vs excitation frequency with varying the UCA's initial radius. Here $\kappa = 1.07$, $\kappa^s = 5 \times 10^{-9}$ N·s/m, $E^s = 0.1$ N/m.

### D. Influence of overpressure

The dependence of the subharmonic component of an UCA on the ambient pressure has been demonstrated as an excellent indicator of the hydrostatic pressure variation (Shi et al., 1999). Inspecting the subharmonic amplitude of an UCA thus could be an effective and noninvasive diagnostic method to estimate the overpressure in heart cavities and major vessels (Dave et al., 2011). Since the onset of the subharmonic oscillation occurs earlier than the growth, saturation, and disappearance of the subharmonic component, studying the dependence of the subharmonic threshold on the overpressure may help better understand the evolution of subharmonic signals. A surface plot of the subharmonic threshold is presented in ***Figure 6***. Here the overpressure $P_{ov}$ is defined as the part exceeding the hydrostatic pressure $p_0$. The cross-section at each $P_{ov}/p_0$ maintains the V-shape, of which the minimum value locates near $f = 2f_0$ (indicated by the blue color). With increasing the overpressure, the threshold increases progressively as well, but the differences between the higher and lower overpressures for any frequency are not drastically changed. It indicates that the overpressure has a mild impact on the subharmonic threshold. This characteristic could be considered when designing UCAs for clinic uses.

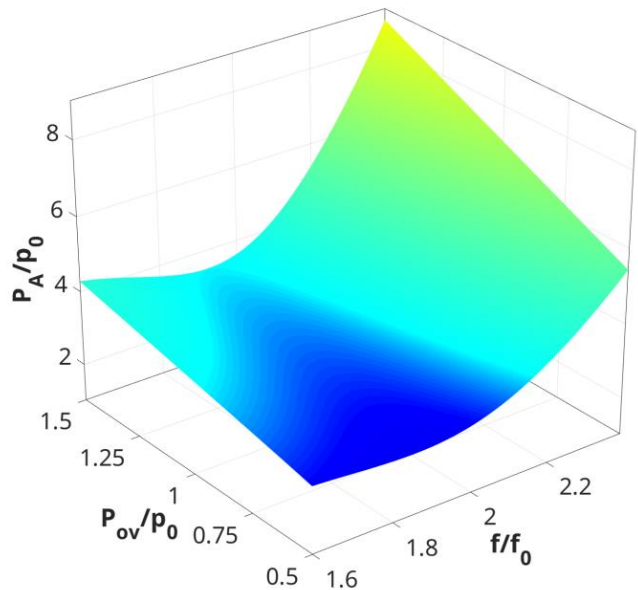

**Figure 6**: Surface plot showing the subharmonic threshold as a function of overpressure and excitation frequency. Here $R_0 = 2.5 \times 10^{-6}$ m, $\kappa^s = 5 \times 10^{-9}$ N·s/m, $E^s = 0.1$ N/m.

### E. Shift of the minimum subharmonic threshold

Some perturbative analyses predicted the minimum subharmonic threshold of a bubble or an UCA occurs when the excitation frequency is twice the linear resonance frequency (Eller & Flynn, 1969; Prosperetti, 1977; Shankar et al., 1999), while other analytical results of an UCA showed that the minimum subharmonic threshold occurred at the frequency less than twice the resonance (Kimmel et al., 2007; Prosperetti, 2013). Additionally, numerical simulations demonstrated that the minimum values of the subharmonic threshold could shift around $f = 2f_0$ (Jimenez-Fernandez, 2018; Katiyar & Sarkar, 2012). Although the numerical investigations suggested that the shift of minimum subharmonic threshold is closely related to the characteristics of the employed shell models, a consistent interpretation of the phenomenon has not been achieved. The analytical results in this article, however, may provide an unambiguous explanation for the above shift.

First, we take a linear undamped system $\ddot{x} + \omega_0^2 x = a\exp(2i\omega_0 t)$ for example, it always has a trivial subharmonic component in the solution $x = -a\exp(2i\omega_0 t)/(3\omega_0^2) + b\exp(i\omega_0 t)$ as the excitation frequency in the force term is twice the resonance frequency $\omega_0$. Then Let's see an undamped nonlinear system, e.g., $\ddot{x} + \omega_0^2 x + x^3 = a\exp(2i\omega t)$, which can be approximated to the linear equation when the cubic term is neglected, with the linear resonance being $\omega_0$. The approximate linear solution implies unbound behavior with increasing the excitation amplitude *a*. However, the nonlinear system does not exhibit unbounded behavior due to the nonlinearity of the cubic term because the amplitude of oscillation will detune the period of oscillation, decreasing its tendency to produce large oscillations. In other words, the resonance frequency of the nonlinear system will depend on the excitation amplitude. For instance, the resonance frequency of this nonlinear system can be approximated by $\Omega = \omega_0 + \varepsilon a + O(\varepsilon^2)$. Thus, for the subharmonic to be triggered, the excitation frequency should approach $2\Omega$, rather than twice the linear resonance $2\omega_0$. Since the resonance frequency of the unperturbed system is the backbone curve, about which the nonlinear resonance frequency of the perturbed system exists (Cenedese & Haller, 2020), we may treat this nonlinear resonance frequency as the resonance frequency of the corresponding perturbed system by assuming the dissipation sufficiently small. This is a major presumption used for carrying out the phase plane analysis. Equation (14) explicitly shows that the subharmonic orbit exists only when the excitation frequency is twice the resonance frequency of the perturbed system given by Eq.(8), in which the resonance explicitly depends on the amplitude of the oscillation and shifts around the linear resonance $\omega_0$ depending on the type of the nonlinearity of the restoring force in Eq.(2). Thus, the excitation-dependent nonlinear resonance contributes to the shift of the subharmonic resonance of the UCA.

Second, for an unperturbed (damped and with external force) system when the dissipation is added, such as the UCA investigated here, the linear resonance frequency becomes the so-called damped resonance $\omega_0\sqrt{1 - \delta_0^2/2}$. It implies the subharmonic resonance should also change accordingly. To see this, we may denote the second modulus of the numerator in Eq.(22) as $W_1$ and the modulus of the denominator as $W_2$. Then in the frequency range of $1.6f_0$ to $2.4f_0$ we see that the ratio is $W_1/W_2 \approx 0.2$, which is almost constant when the linear damping is kept below 0.3. Thus, by checking the first modulus of the numerator in Eq.(22), we readily find that the minimum occurs when the excitation frequency approaches to $2\omega_0\sqrt{1 - \delta_0^2/2}$, which is twice the damped resonance frequency of the UCA. This partly explains some numerical and analytical results demonstrating that the minimum subharmonic threshold of an UCA could be less than twice the linear resonance frequency of the UCA (Jimenez-Fernandez, 2018; Katiyar & Sarkar, 2012; Kimmel et al., 2007). Noticing that when the damping is too large or the excitation frequency is too far away from twice the linear resonance, $W_1/W_2$ cannot be approximated by a constant. In this situation, the above linear approximation will not be accurate, and the shift of the minimum subharmonic threshold will also depend on $W_1/W_2$ (a result of nonlinear interactions).

From the above analysis we can conclude that the shift of the minimum subharmonic threshold of the UCA is the result of the excitation-dependent nonlinear resonance and the introduced damping, as well as their interactions in case the system is highly nonlinear. The damping alone is not sufficient ton explain the shift, and vice versa. (Katiyar & Sarkar, 2012) also claimed the shifting of subharmonic was not caused mainly by the damping based on their calculations for scattering pressures, rather than the bubble radius investigated here.

**F. Subharmonic threshold at low excitation pressure**

The subharmonic threshold of an UCA subjected to single-frequency excitation could be significantly lower than theoretical predictions (near $f = 2f_0$) in the perturbative analysis (Lotsberg et al., 1996; Shankar et al., 1999), and a potential cause was thought to be the nonlinear shell behaviors (Jimenez-Fernandez, 2018; Prosperetti, 2013; Sijl et al., 2010). Although nonlinear shell behaviors might be a cause, the discussion of the nonlinear dilatational viscosity--that is another important parameter also contributing to the nonlinear behaviors of the shell--was often absent in those works. Other experimental observations and numerical simulations also showed that the minimum subharmonic threshold could further shift to the main resonance frequency of an UCA (Chomas et al., 2002; Faez et al., 2012). This was even more difficulties to predict theoretically. This is because the subharmonic threshold that was derived from those classical perturbation analyses only employed an approximated solution near the subharmonic resonance $f = 2f_0$. It is not valid in predicting the threshold when the excitation frequency is too much far away from the subharmonic resonance. However, this is not the case in the present study because, in deriving Eq.(22), we did not restrict the solutions in the neighborhood of $f = 2f_0$. Therefore, we may use this equation to qualitatively study the subharmonic threshold in a broader range of excitation frequencies, for instance, including the linear resonance. *Figure 7* shows two curves with different dilatational viscosities, of which the solid one is calculated with a lower dilatational viscosity. It exhibits two minima, one is near the main resonance, and the other is near the subharmonic resonance. The first minimum is lower than the second one. When the dilatational viscosity is increased, the dashed curve shows that these two minima have merged into a single minimum that locates near the linear resonance, at which the threshold value is also less than that of at $f = 2f_0$. These two curves greatly resemble those obtained in the numerical simulations (Jimenez-Fernandez, 2018; Katiyar & Sarkar, 2012). Here the results from the lowest order approximation with linear shell behaviors again suggest that the dissipation and shell elasticity interact nonlinearly and conspire to shift the minimum subharmonic threshold.

We remark that the accuracy of the above results is mainly determined by truncations of the Fourier series of the subharmonic-free solution. Additionally, the amplitude-dependent nonlinear resonance cannot be captured by Eq.(22). To include the effect, we may need to employ the multiscale or Lindstedt- Poincaré method, but the result will lose its simplicity and explicitness.

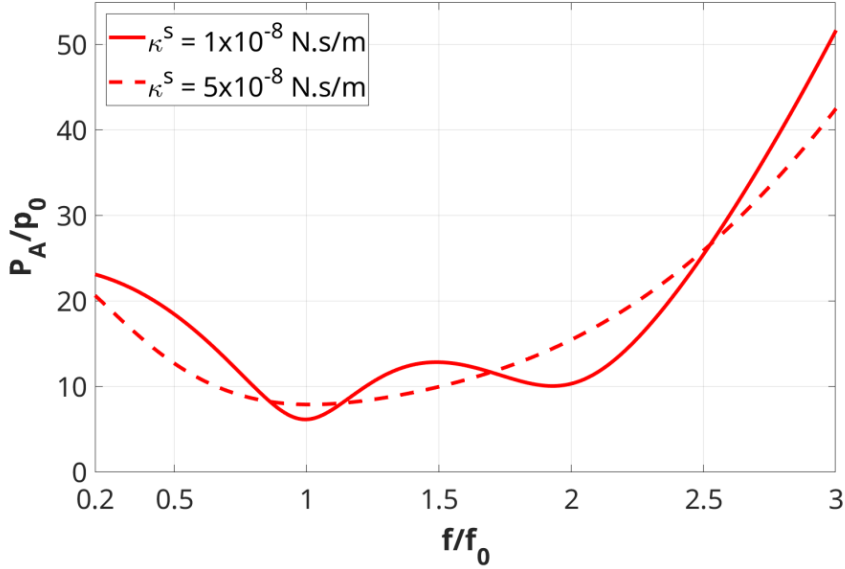

**Figure 7**: The subharmonic threshold calculated from Eq.(22) in a wider frequency range. The dilatational elasticity is fixed at 0.5 N/m for both curves. Here $R_0 = 2.5 \times 10^{-6}$ m.

### G. Concerns on nonspherical oscillations

Besides the spherical (or volume mode) oscillation, the translational (or dipole) motion and nonspherical (or shape mode) oscillations of an UCA may coexist when it is subjected to an oscillating force. These types of dynamics can couple with each other, possibly affecting the subharmonic oscillation of the UCA. As the translational motion is triggered by the interaction of neighboring shape mode oscillations (translation due to the primary Bjerknes is neglected here), we then need only to investigate the influence of nonspherical oscillations on the subharmonic oscillations.

However, the impact of nonspherical oscillations on the subharmonic threshold of the UCA is not readily inferred because some of the shape modes also generate the subharmonic component even though it decays rapidly. This subharmonic signal due to nonspherical oscillations could mix with the one generated by the spherical oscillation, bringing several difficulties in identifying the source of subharmonic emissions from the UCA. Fortunately, like the threshold phenomenon of subharmonic emission due to the spherical oscillation, the onset of shape mode oscillations also requires excitation pressures above certain values, below which the shape mode oscillations would not start. If the excitation pressures exceed the threshold of shape mode oscillations, the UCA will have nonspherical oscillations and potentially generate subharmonic components. Therefore, by investigating the two thresholds we may be able to differentiate the subharmonic signals and identify the source of subharmonic emissions. Within the current framework, by also ignoring the higher-order effects, e.g., mode interactions and the thickness of the boundary layer, we may write the threshold for shape oscillations in the form of (Guédra & Inserra, 2018)

$$P_{A,n} = 2\rho R_0^2 \frac{\left|-\frac{1}{4}\omega^2 + \frac{1}{2}\mathbf{i}\delta_{on}\omega\omega_{on} + \omega_{on}^2\right| \cdot \left|-\omega^2 + \mathbf{i}\omega\omega_0\delta_0 + \omega_0^2\right|}{\left|3\omega_{on}^2 - (n-1)\omega_0^2 - (n-1)\left(-\omega^2 + \mathbf{i}\omega\omega_0\delta_0 + \omega_0^2\right)\right|} \qquad (23)$$

in which $n \geq 2$ is the order of shape mode, and other coefficients are given by (Liu & Wang, 2016; Loughran, Eckersley, & Tang, 2012)

$$\omega_{0n}^2 = (n-1)(n+1)(n+2)\frac{1}{\rho R_0^3}\left(\gamma_0 + \frac{2E^s}{2n^2 + 2n - 1}\right)$$

$$\delta_{0n} = (n+2)\frac{1}{\rho R_0^2 \omega_{0n}}\left[2(2n+1)\mu + (n^2 + 4n + 2)\frac{\kappa^s}{R_0}\right] \qquad (24)$$

Note that in deriving the above equations, we have neglected the radiation damping due to the compressibility of the water, which will increase the threshold $P_{A,n}$. **Figure 8**, shown as an example, demonstrates that the subharmonic threshold calculated from Eq.(22) is always lower than the threshold of shape instability evaluated from Eq.(23) when the excitation frequencies are near twice the resonance. Thus, we can conclude that the subharmonic component due to spherical oscillation is easier to be triggered than those of potentially due to nonspherical oscillations if the excitation frequencies are not far away from $f = 2f_0$. However, outside this region, the threshold of shape oscillation could be lower than the subharmonic threshold. In this case, the onset of shape oscillation will affect the amplitude of volume mode oscillation, of which the first-order approximation may be written by

$$A_s = \frac{P_A}{\rho R_0^2}\frac{1}{-\omega^2 + \mathbf{i}\omega\omega_0\delta_0 + \omega_0^2}\left[1 + \frac{1}{P_A}\frac{(4n+9)\omega_{on}^2 - 2(n+1)\omega_0^2}{4(2n+1)(n+1)}S_n^2\right] \qquad (25)$$

where $S_n$ is the complex amplitude of shape oscillations (Guédra & Inserra, 2018). Replacing $A$ in Eq.(20) by $A_s$ could offer a means to evaluate to the subharmonic threshold considering the effect of nonspherical oscillations. Note that the higher excitation pressures required by the shape mode oscillations will also generate higher harmonic components of large amplitudes, which cannot be captured by the current first-order approximation and is left for future exploration.

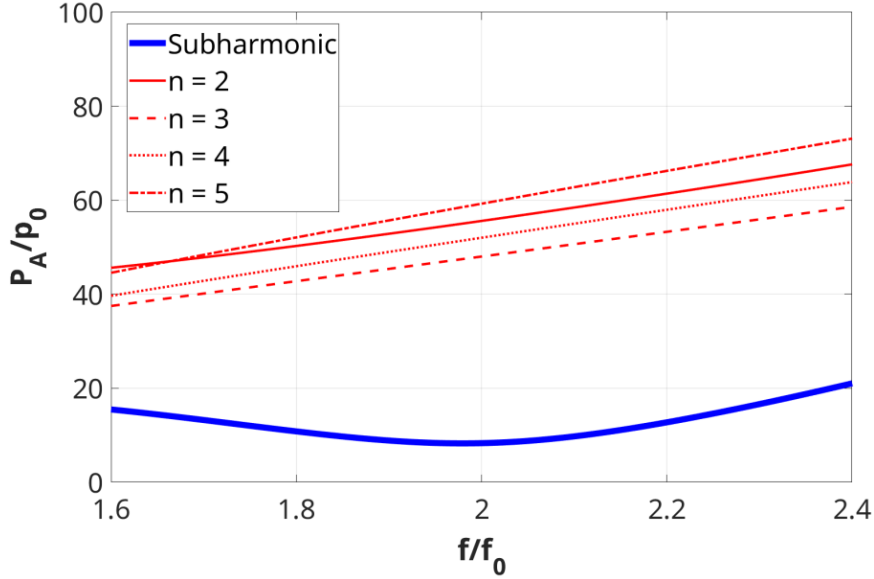

**Figure 8**: The thresholds for the shape instability of mode n = 2 to 5 (red thin curves) vs the subharmonic threshold (thick blue curve). Here $R_0 = 2.5 \times 10^{-6}$ m, $\kappa^s = 5 \times 10^{-9}$ N·s/m, $E^s = 0.5$ N/m.

## IV. CONCLUSION

We have presented a framework to analyze the thresholds for the subharmonic existence and subharmonic instability of an UCA subjected to single-frequency excitation. We first studied the unperturbed system (the governing equation without damping and external force) of the UCA in the phase plane and obtained an exact solution that was used to calculate the nonlinear resonance of the corresponding perturbed equation (with damping and external force) of the UCA when the effects of dissipation and external forcing were not large. By investigating the energy balance during UCA oscillation, we have derived Eq.(15) that is equivalent to the results of the classic subharmonic Melnikov's function and capable of predicting the threshold for the existence of the subharmonic orbit. Most importantly, we sketched an idea to study the threshold for the subharmonic instability by separating the subharmonic-free and subharmonic oscillations in the full nonlinear governing equation. By using only the first-order approximation of the subharmonic-free oscillation, we have obtained Eq.(22) that well predicted a set of classic numerical results of free bubbles. It also matched well with the numerical calculated subharmonic threshold of an UCA near twice the linear resonance. We then investigated the influence of shell parameters on the threshold for subharmonic instability and found both of dilatational viscosity and elasticity promoted the threshold when they increased. In contrast, the increase of polytropic constant and UCA's initial radius decreased the subharmonic threshold. Based on these analytical results, we demonstrated that the shift of subharmonic resonance could be due to the amplitude-dependent nonlinear resonance, introduced damping, and their nonlinear interactions. Eq.(22) qualitatively characterized the experimentally observed threshold phenomena, such as the existence of the minimum subharmonic threshold at the linear resonance. Finally, we briefly studied the threshold for shape mode oscillations of the UCA, demonstrating that the excitation pressures for the onset of subharmonic oscillation are much lower than those of shape mode oscillations when the excitation frequencies were near twice the linear resonance.

**ACKNOWLEDGMENTS**

I appreciate the 4 anonymous reviewers' suggestions form JASA, especially for the 3 positive comments.